# A Bulk-Controlled Low-Voltage CMOS Quadrature Oscillator

R. Picos, M. A. Calafat, K. Suenaga, S. Bota, M. Roca, E. Isern, E. García-Moreno

Grup de Tecnologia Electrònica, Universitat de les Illes Balears, rodrigo.picos@uib.es, +34-971-173-227

*Abstract*— **In this paper, an schema for controlling the oscillation frequency of a quadrature oscillator is proposed. The method involves controlling the threshold voltage of the PMOS transistors in the inverter through control of the bulk bias voltage. Results obtained using HSPICE simulation are presented in a technology of 0.35μm, and experimental results using discrete elements (HEF4007) are also shown. Both sets of experiments show the effectiveness of the technique.**

**Index Terms— MOS circuits, quadrature oscillator, bulk voltage control.**

## I. Introduction

THE rapidly growth of the wireless communication field has its origin in the existence of low-cost solutions for radio-frequency subsystems [1-3]. Among other major design issues, portability is perhaps one of the more demanding ones. Portability requires not only high integration levels in order to provide the expected processing power, but also very low power consumption. This need is even greater when mixed-signal or analog circuits are involved. Innovative solutions are thus needed to satisfy the market demand.

A first indication on how to get low-power circuits comes from a detailed examination of the following very simple expression to estimate power consumption:

$$P = f \cdot C_L \cdot V_{DD}^2 \qquad (1)$$

where $f$ is the switching frequency, $C_L$ is the equivalent capacitive loading at the output of the circuit, and $V_{DD}$ is the power supply voltage. This relation implies that power dissipation reduction can be achieved by simply cutting down $V_{DD}$. However, this reduction in power consumption is not so clear when the CMOS operates at very low voltages, below the threshold voltages of MOSFETs, or in the range of few hundreds of millivolts. At these very low supply voltages, the implicit linearity in the above equation is no longer true, and the current driving capability of MOSFETs decreases exponentially, while the capacitances in the circuits are to remain the same, or even increase, thus slowing down the circuits or even causing them to fail to operate.

In RF circuits the oscillator is a building block that is both high-consuming and important [4,5]. In this work, we will propose a method to control a VCO in very low operating voltages. Our proposal involves using the PMOS as what they really are: four terminals devices, where we will use the bulk terminal as the frequency control terminal, instead of connecting it to $V_{DD}$.

The paper is organised in five sections. After this introduction, Section II describes the basic oscillator, without any modifications. Section III presents the required modifications and simulations showing the applicability of the technique. Section IV presents experimental results obtained using HEF4007 discrete elements. And, finally, Section VI details the main conclusions of the paper.

## II. Basic Oscillator

The CMOS ring oscillator schema of Fig. 1a has been chosen to implement our VCO. The same control technique can be applied to other structures. Each inverter stage corresponds to the static CMOS inverter of Fig. 1b. The most basic ring oscillator employs an odd number of stages as delay cells, unfortunately quadrature outputs require a ring with even number of stages, therefore to achieve oscillation feed forward inverters have to be added between nodes with opposite-phase signals [4]. In contrast to a ring with an odd number of inverters, now two transitions are travelling through the ring. As there is a rising and failing transition at any time the single-ended quadrature ring oscillator draws a nearly constant supply current and minimises switching noise on the supply lines.

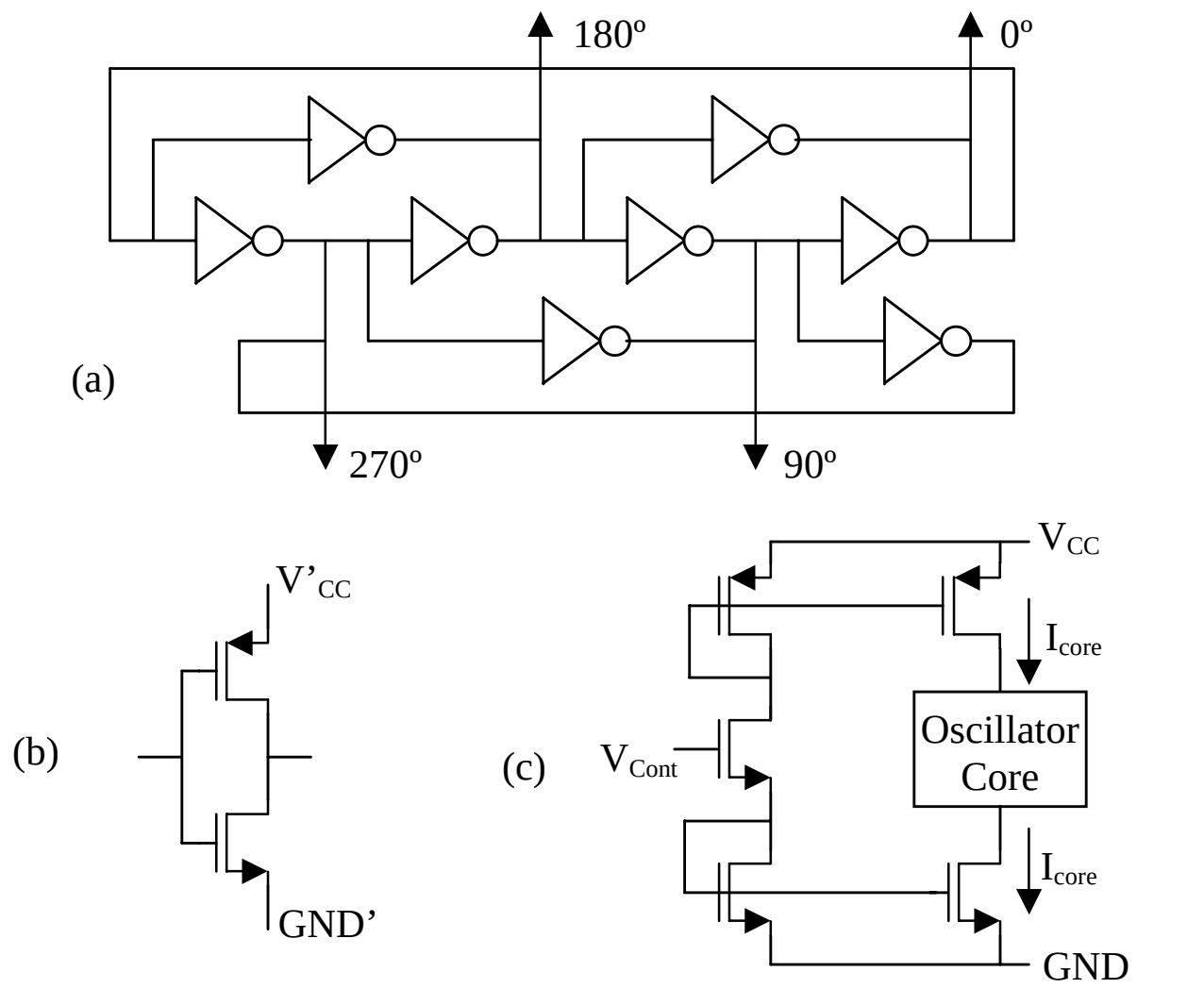

Fig. 1. (a) Core of the quadrature four-stage ring oscillator. (b) CMOS inverter. (c) Basic frequency control.

Frequency control can be achieved by programming the current through individual inverter stages, but it is preferred to control the total quadrature oscillator core current as proposed in [5]. This is done with a PMOS current source connected to the positive supply and a NMOS current sink connected to the negative supply (Fig. 1c). The core current is controlled by applying appropriate voltage to $V_{cont}$. With this approach, the optimum strength ratio between main and feedforward inverters is maintained at any core current. Moreover, the current source transistors offer supply rejection and attenuate the switching noise.

This VCO has been designed in a 0.35 µm CMOS process. The outputs, simulated with HSPICE [6], are plotted in Fig. 2 at $V_{DD}$ = 3 V and $V_{cont}$ = 2.09 V which sets the oscillation frequency at 240 MHz. It can be seen that they are roughly sinusoidal. Fig. 3 shows the oscillation frequency and the average current consumption versus the control voltage.

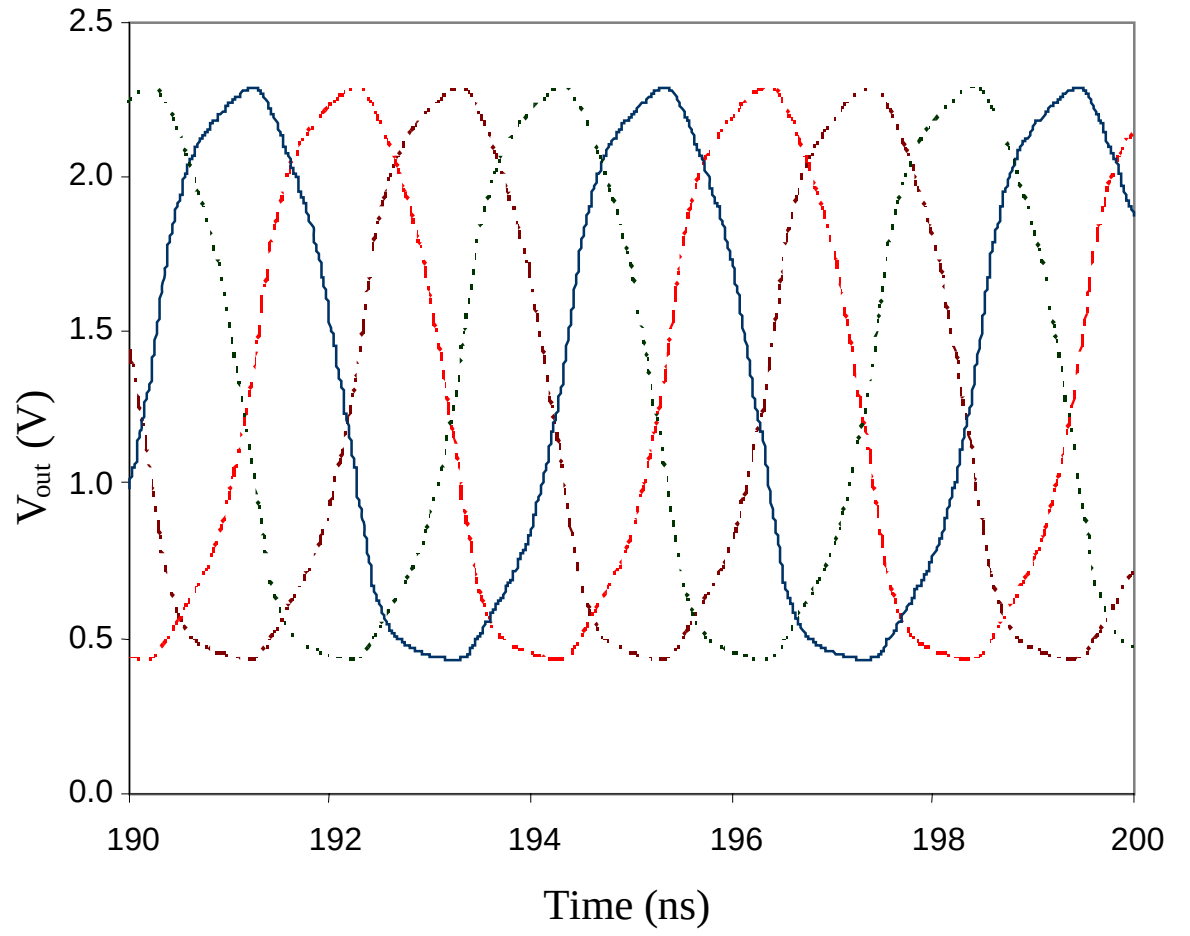

Fig. 2. Simulated outputs of the VCO. It can be seen that the outputs are in quadrature, and that they are roughly sinus-like. Oscillation frequency is around 240 MHz.

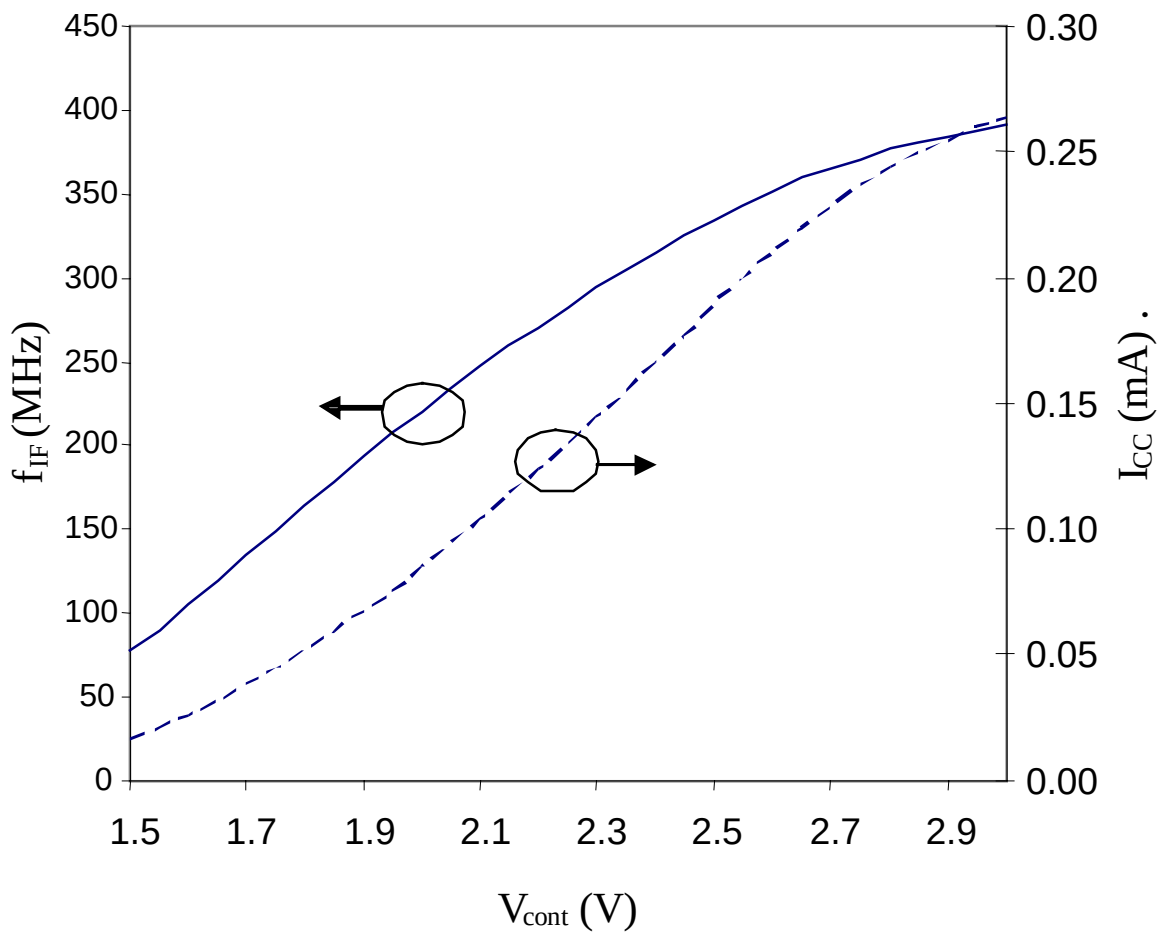

Fig. 3. Oscillation frequency (–) and average current consumption (--) of the VCO as a function of control voltage.

## III. Proposed Oscillator

The control strategy proposed in the previous section, though effective, can be problematic in very low voltage circuits. In this work, we propose controlling the oscillator by modifying the threshold voltage of the transistors, in a similar way to that proposed in [7] or in [8], where they both use ring oscillators. It is a fact that the oscillating frequency of the VCO depends on the threshold voltage ($V_T$) of the transistors, for a given value of the power supply voltage $V_{DD}$. This $V_T$ can expressed as:

$$V_T = V_{T0} + \gamma \sqrt{2V_{Th} \ln\left( N_B / n_i \right)} - V_B \qquad (2)$$

where $V_{T0}$ is the unbiased threshold voltage, $N_B$ is the doping density, $n_i$ is the intrinsic doping density, $V_{Th}$ is the thermal voltage ($k_BT/q$), $\gamma$ is the bulk factor ( $\gamma = \sqrt{2\varepsilon_{Si} q N_B} / C'_{ox}$ ). From the above formula, we can see that if we vary the bulk bias voltage ($V_B$), the threshold voltage will change accordingly. In the technology we are using (a N-well 0.35 µm CMOS process), it is best advised not to tamper with the NMOS bulk voltage, because we could interact with other parts of the system though the common bulk. On the other hand, since the PMOS transistors are to be isolated in their own N-well, it is safe to change their bulk voltage, provided that we don't go below the ground voltage nor over $V_{DD}$. Moreover, since the $V_T$ variation depends on the bulk doping $N_B$, the change in the PMOS threshold will be greater than that of the NMOS, for the same bulk voltage variation. In a dual-well process, we would be able to change both voltages, and thus achieve a greater control range.

The schematics of the proposed oscillator is the same than the one shown in Fig. 1.a, whith the only difference being that the inverters are the ones shown in Fig. 4, where the control terminal is the bulk voltage of the PMOS transistors, which is common to all of them, since they are all in the same N-Well.

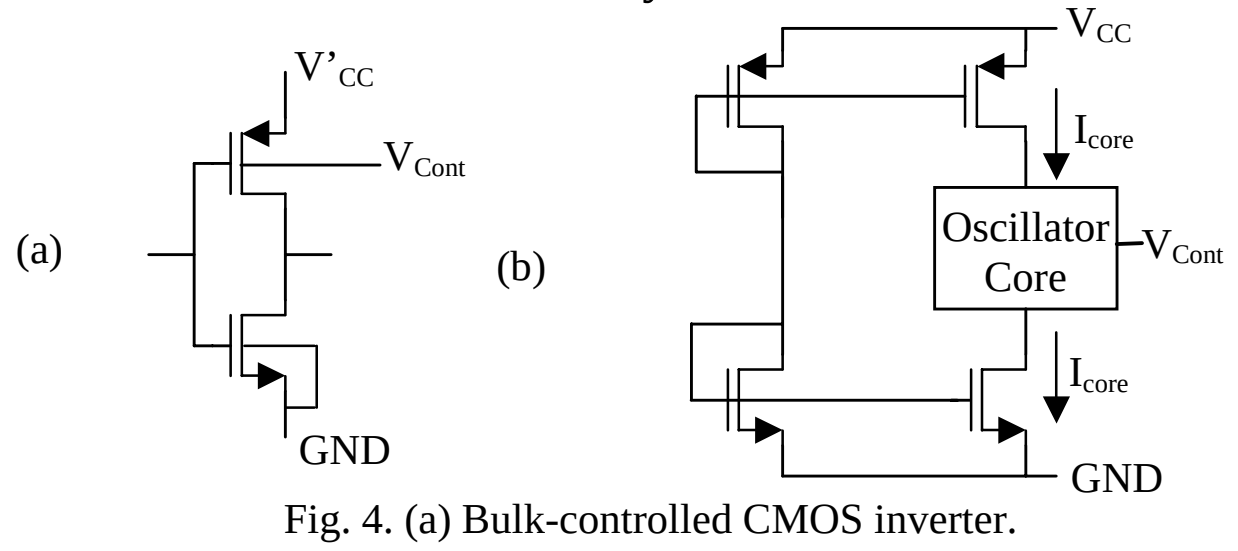

Fig. 4. (a) Bulk-controlled CMOS inverter.

This VCO has also been designed in the same 0.35 µm CMOS process than the previous one. The circuit has been simulated with HSPICE [6] at $V_{DD}$ = 3 V and $V_{cont}$ = 3 V, resulting in an oscillation frequency of 540 MHz. The outputs are still roughly sinusoidal, as before. Figure 5 shows the oscillation frequency versus the control voltage for $V_{DD}$ = 3 V. It can be seen that there is a maximum frequency of around 547 MHz at $V_{Cont}$ a few tenths of volts below $V_{DD}$.

However, the best application of this oscillator is to be found when $V_{DD}$ is comparable to the threshold voltage. This is because then the impact of the variations of $V_T$ will be greater compared to the range of operation of the inputs of the transistors (rail to rail in the better case). These threshold voltages (unbiased) are around 0.4 V and –0.45 V for the NMOS and the PMOS, respectively. We simulated the variation of the frequency versus the control voltage for various $V_{DD}$ between 1 V and 0.6 V. We see (Fig. 5) that the range of variation of the frequency is between 68MHz and 42MHz for the 1V and between 690KHz and 290KHz for the 0.6 V case. As expected, the relative variation is greater in the case with the lower supply voltage, though the frequency is lower.

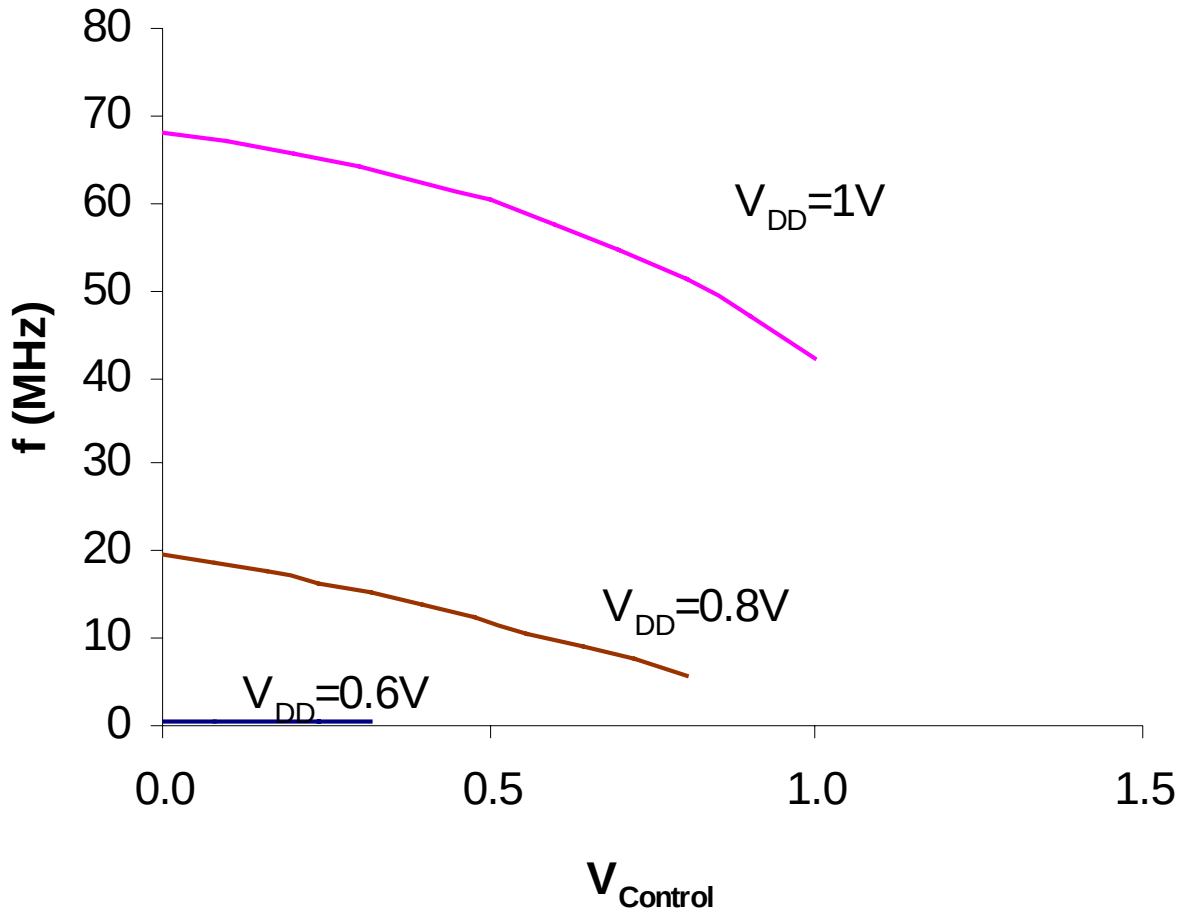

Fig. 5. Simulated Oscillation frequency and average current consumption of the VCO in 0.35 μm technology as a function of the control voltage, for $V_{DD}$=3.3V.

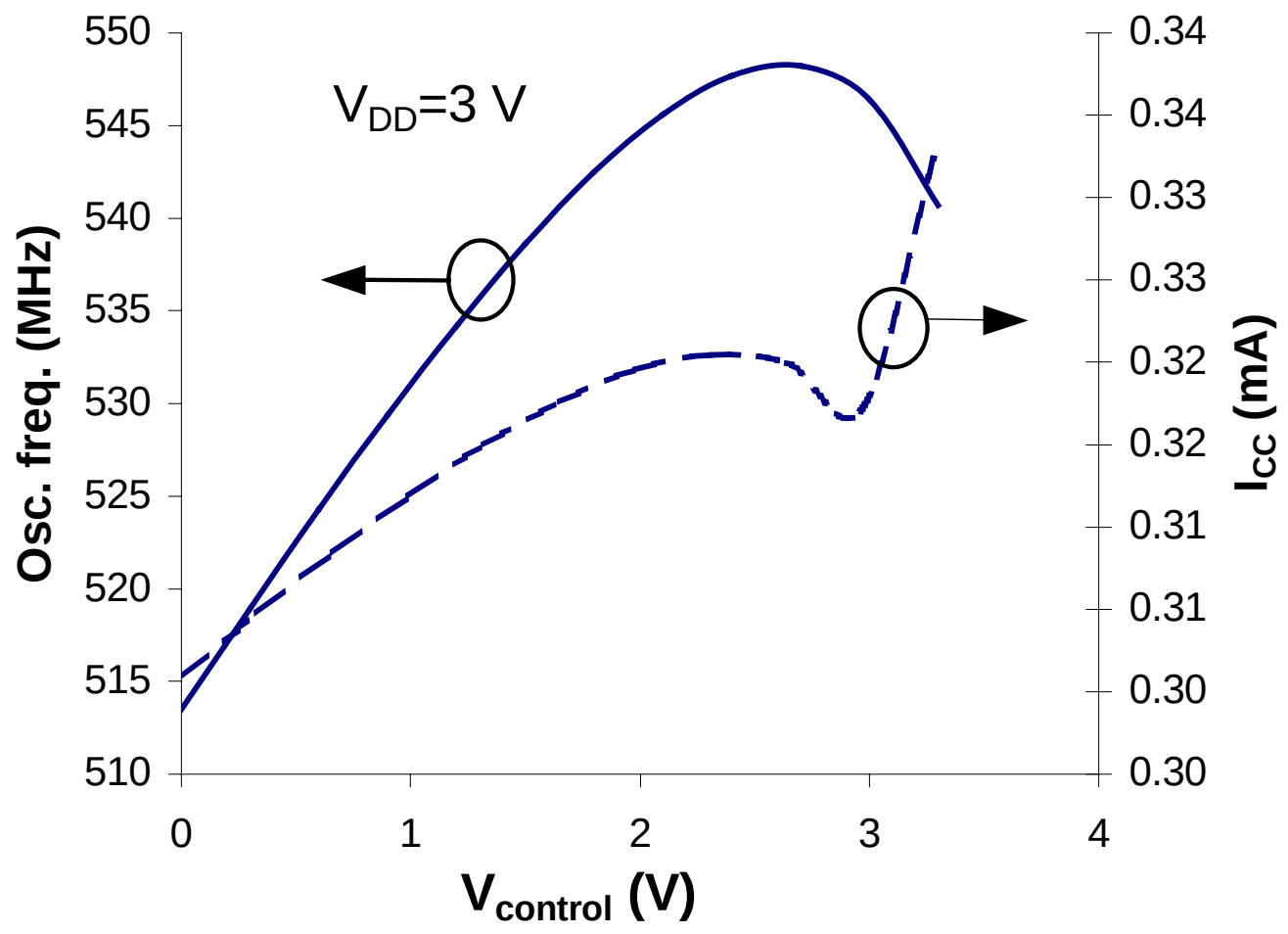

Fig. 6. Simulated Oscillation frequency and average current consumption of the VCO in 0.35 μm technology as a function of the control voltage, for $V_{DD}$=3.3V.

There is a point worth some comments. If we look at Fig. 6, where the frequency vs. the control voltage has been plotted for a normal voltage ($V_{DD}$=3 V), we can see that there is a maximum. That is, the frequency increases with the control voltage, until a maximum is reached. Then, the frequency begins to fall down. The position of this maximum has been plotted as a function of the supply voltage in Fig. 7. In this figure, we can see that the position of this maximum tends to shift to lower voltages as we reduce the supply voltage. In the limit, this means that, when reducing the supply voltage, we reach a point (in the case of our technology, around 1.3V) whereupon the maximum frequency voltage is at a control voltage of zero, as can be seen in Fig. 5.

This behavior is due to the fact that the position of the maximum is related to the point where the parasite diode turns on. This point is reached when the argument of the square root in equation 2 (that can be related to the doping) becomes negative. Thus, trying to increase the threshold voltage too much (that is, reducing the control voltage), implies a DC current injection through the bulk, thus reducing the oscillation frequency.

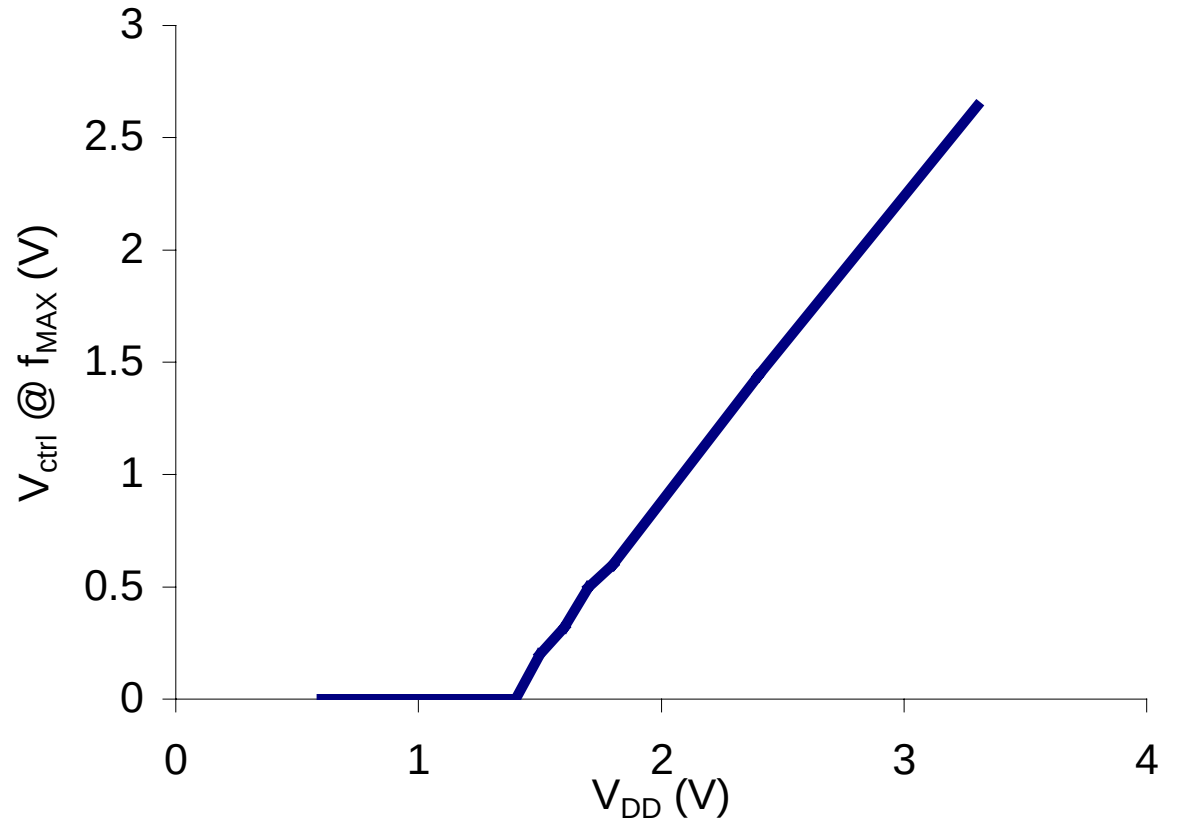

Fig. 7. Control voltage providing the maximum frequency as a function of supply voltage for the VCO implemented in a 0.35 μm technology.

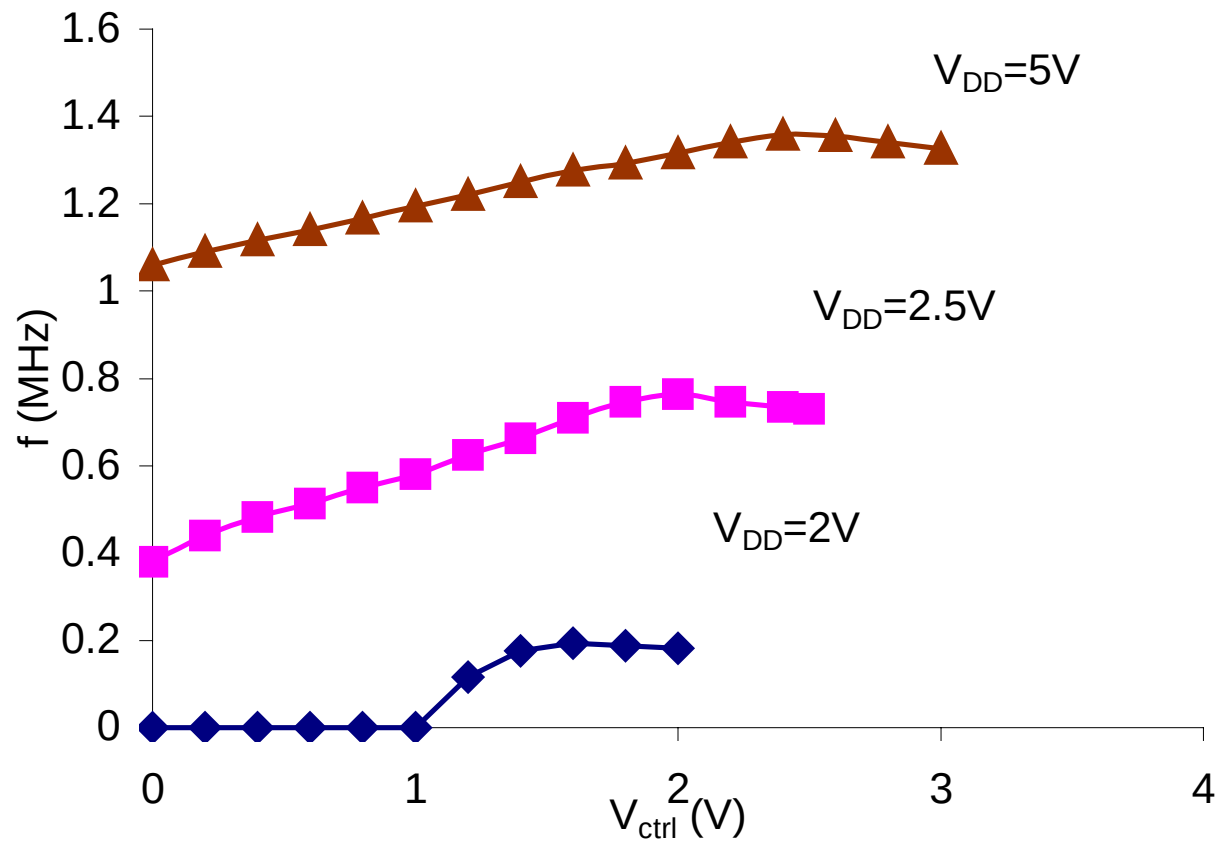

Fig. 8. Experimental oscillation frequency of the VCO with HEF4007 as a function of control voltage, for different power supply voltages.

## IV. Experimental Results

In order to validate the results obtained in previous sections, we have implemented the VCO using discrete elements. Since we did not had access to 0.35μm inverters (they have been sent to the foundry, but they haven't yet been received back), we used the classical HEF4007 CMOS package of PMOS and NMOS transistors. Using this elements, we put together the proposed oscillator. In this case, though we actually were able to change the bulk voltage of the NMOS transistors, we did maintain it constant in order to keep the same conditions than before. In this technology, the threshold voltages are around 1.4V for the NMOS and –1.4V for the PMOS transistors.

Recommended supply voltages for this technology are between 5V and 12V. However, we ran the oscillator at several different $V_{DD}$ voltages, showing the effectiveness of the bulk control technique, even at voltages as low as 2 V, that are well below the addition of the PMOS and NMOS voltages. This way, operation in the weak inversion regime has been demonstrated. It has to be noted also that the maximum of the oscillation frequency still appears in this technology, thus removing the possibility of a numerical artefact due to the simulator.

Figure 8 shows the frequency versus the control voltage for three different supply voltages of 2 V, 2.5 V and 5 V. It can be seen that, as expected, the greater relative range of variation (between 0.4MHz and 0.8MHz) was found in the 2.5V voltage. The upper voltage (5V) provided a variation range of 1.05MHz to 1.38MHz. Further experiments conducted with greater supply voltages showed that no significant variation of the frequency was achieved. This is so because the relative variations of the threshold voltage are too small to affect the behavior of the oscillator at high voltages, as explained before.

## V. Conclusions

A quadrature VCO has been proposed. This oscillator uses the bulk voltage as the control voltage, thus making it able to function at lower voltages than other techniques. A margin of frequency variation of 515MHz to 540MHz has been achieved under normal operating conditions for a 0.35mm technology. If we reduce the supply voltage, then the relative variation range increases, even if the frequency decreases. For instance, for a supply voltage of 0.6V, an oscillation range from 290KHz to 690KHz has been observed.

In order to test the procedure, the oscillator has been implemented using discrete elements (HEF4007). Experimental results show that the technique is well suited to control the oscillation at supply voltages well below the nominal ones, when the transistors are forced to operate in the weak inversion region.

## Acknowledgement

This work has been partially supported by the Spanish TEC2006-04103 project.